# Superconducting epitaxial YBa$_2$Cu$_3$O$_{7-\delta}$ on SrTiO$_3$ buffered Si (001)

K. Ahmadi-Majlan[a], H. Zhang[b], X. Shen[c,d], M.J. Moghadam[a], M. Chrysler[a], P. Conlin[a], R. Hensley[a], D. Su[e], J. Y. T. Wei[b], and J. H. Ngai[a*]

[a]Department of Physics, University of Texas at Arlington, Arlington, TX 76019, USA
TX 76019, USA
[b]Department of Physics, University of Toronto, Toronto, ON M5S 1A7, Canada
[c]Center for Functional Nanomaterials,
Brookhaven National Laboratory, Upton, New York 11973, USA
[d]National Laboratory of Solid State Microstructures and Department of Materials Science and Engineering, Nanjing University, Nanjing 210093, P. R. China

*email: jngai@uta.edu

**Abstract.** Thin films of optimally-doped (001)-oriented YBa$_2$Cu$_3$O$_{7-\delta}$ are epitaxially integrated on silicon (001) through growth on a single crystalline SrTiO$_3$ buffer. The former is grown using pulsed-laser deposition and the latter is grown on Si using oxide molecular beam epitaxy. The single crystal nature of the SrTiO$_3$ buffer enables high quality YBa$_2$Cu$_3$O$_{7-\delta}$ films exhibiting high transition temperatures to be integrated on Si. For a 30 nm thick SrTiO$_3$ buffer, 50 nm thick YBa$_2$Cu$_3$O$_{7-\delta}$ films that exhibit a transition temperature of ~ 93 K, and a narrow transition width (< 5 K) are achieved. The integration of single crystalline YBa$_2$Cu$_3$O$_{7-\delta}$ on Si (001) paves the way for the potential exploration of cuprate materials in a variety of applications.

1. Introduction

Complex transition-metal oxides exhibit a range of material behaviors that can potentially be exploited in future microelectronic, sensing and energy harvesting technologies. Amongst complex oxides, high-temperature superconducting cuprates are of fundamental and technological importance. Efforts to elucidate the microscopic mechanism governing superconductivity in the cuprates have improved our fundamental understanding of strongly correlated phenomena in condensed matter systems [1]. In regards to technological importance, cuprate materials such as $YBa_2Cu_3O_{7-\delta}$ (YBCO), have been explored for potential use in applications such as infrared detectors, microwave filters and superconducting quantum interface devices (SQUID) [2-8]. For many applications, high transition temperatures ($T_c$) and narrow transition widths are desired, and thus single crystalline films are preferred [9-10] .

However, the growth of single crystalline YBCO films for potential applications is challenging for two reasons. First, epitaxial YBCO films are typically grown on single crystal oxide substrates that are costly, such as $SrTiO_3$ (STO) and $La_{0.18}Sr_{0.82}Al_{0.59}Ta_{0.41}O_3$ (LSAT). Second, single crystal oxide substrates presently cannot be synthesized in large sizes. YBCO films have been grown on Si using various techniques, such as magnetron sputtering [11-14], and pulsed laser deposition (PLD) [15-21]. Direct deposition of YBCO on Si results in interdiffusion due to the high temperatures that are required [9]. Consequently, the YBCO films deposited directly on Si are polycrystalline in nature and exhibit relatively low transition temperatures and broad transition widths. To minimize interdiffusion, intermediary buffer layers such as $ZrO_2$, MgO, AlN, etc. have been introduced between YBCO and Si [9,14,21-28]. Such buffer layers significantly improved superconducting properties, yielding $T_c$'s from 55 K to 88 K for 160-400 nm thick

YBCO films deposited on top. The quality of the buffer between YBCO and Si is thus critical to mitigating interdiffusion and improving film quality and superconducting properties.

Here we utilize a single crystalline STO buffer to epitaxially integrate YBCO on Si (001). The former is grown on Si using oxide molecular beam epitaxy (MBE), and the latter is grown using pulsed laser deposition (PLD) shown schematically in Figure 1. X-ray diffraction (XRD) and cross-sectional high resolution transmission electron microscopy (HRTEM) and scanning transmission electron microscopy (STEM) confirm the single crystalline nature of the YBCO and STO buffer. Our 50 nm thick YBCO films on Si exhibit high transition temperatures (93 K), and narrow transition widths (< 5 K), which can be attributed largely to the single crystalline nature of the STO buffer. The realization of high quality, epitaxial YBCO films on Si opens the pathway to explore YBCO for applications.

## 2. Experimental details

The STO buffer is grown using reactive MBE in a custom-built chamber operating at a base pressure of < 2 x $10^{-10}$ Torr. Undoped, 0.5 mm thick, 2" diameter (001)-oriented Si wafers (resistivity of 14–22Ωcm MTI) are introduced directly into the MBE chamber and exposed for ten minutes at room temperature to activated oxygen, generated by a radio frequency plasma source (Veeco) operating at 220 W in an $O_2$ background pressure of 5 x $10^{-6}$ Torr. The activated oxygen removes residual organic material from the surface of the Si wafer. The native layer of $SiO_x$ is removed through a Sr desorption process that involves the deposition of 2 monolayers (ML) of Sr at a substrate temperature of 560 $^0$C. All source materials are evaporated using thermal effusion cells (VEECO, SVT Associates), and fluxes are measured using a quartz crystal microbalance (Inficon). The silicon wafer is continuously rotated to maintain uniform deposition. The substrate is then heated to 860 $^0$C at which the Sr reacts to form SrO, which readily desorbs from the clean Si surface as shown in Figure 2(a) [28]. The substrate is then cooled to 660 $^0$C at which 0.5 ML of Sr metal is deposited to form the 2 x 1 reconstructed template necessary for subsequent epitaxy of the STO [29]. 3 ML of SrO and 2 ML of TiO$_2$ are co-deposited at room temperature on the 0.5 ML Sr template and then subsequently heated to 500 $^0$C to form 2 unit-cells of crystalline STO. Additional layers of STO are grown at substrate temperature of 625 $^0$C by co-depositing Sr and Ti at a 1:1 ratio in a background O$_2$ pressure of 3 x 10$^{-7}$ Torr. Figure 2(b)(c) and (d) show reflection high energy electron diffraction (RHEED) patterns taken along the [10] and [21] and [11] directions of a 30-nm thick STO film grown on Si. The strong intensity and streaky pattern indicate excellent crystallinity and 2-dimensional growth. Corroborating the RHEED, atomic force microscopy (AFM) (Park XE-70) of the surface of the STO reveals roughness that is < 1 nm, consistent with a 2-dimensional growth mode as shown in Figure 2 (e) and (f).

After growth of STO on Si, the wafers are removed from the MBE chamber and transported in ambient conditions to a separate PLD chamber. Smaller 5 × 5 mm pieces of STO buffered Si are broken off the 2" wafers and glued to a contact heater using Ag paste (SPI products). The substrate is then heated to 850 $^0$C in a background O$_2$ pressure of 200 mTorr, at which the YBCO is deposited by striking a stoichiometric polycrystalline target with a 248-nm excimer laser operating at a rate of 2 Hz. The stoichiometric YBCO target is situated ~ 6 cm from the substrate and is continuously rotated during deposition. After growth, the PLD chamber is back-filled with 1 atm of O$_2$ and the YBCO film is slowly cooled to 300 $^0$C at a rate of 12 $^0$C/min, and then furnace cooled to room temperature.

3. **Results and Discussion**

HRTEM and STEM imaging provide real-space structural characterization of the YBCO/STO/Si heterostructures. The former (latter) is obtained using a JEOL JEM2100F TEM operated at 200 kV (aberration-corrected Hitachi HD2700C dedicated STEM). The cross-sectional specimen along the <100> direction of Si was prepared by focused ion beam using a FEI-Helios system. Figure 3(a) shows a HRTEM of the combined stack comprised of 50 nm of YBCO grown on 30 nm of STO on Si, as indicated. The epitaxial relationship of the combined YBCO/STO/Si stack is YBCO (001) || STO(001) || Si(001) and YBCO(100)/(010) || STO(100) || Si(110). A ~ 5 nm thick layer of $SiO_x$ situated between the Si substrate and STO buffer is also observed. Since the conditions used to grow the epitaxial STO buffer does not create $SiO_x$ at the interface, the latter forms during deposition of the YBCO, which occurs at a much higher temperature and $O_2$ pressure. It should be noted that the STO buffer does not react and intermix with the Si substrate at high temperatures and $O_2$ pressures during the PLD deposition of YBCO, namely, the buffer remains 30 nm thick despite the formation of $SiO_x$. High angle annular dark-field (HAADF) STEM images reveal the *c*-axis oriented nature of the YBCO film and the abruptness of the interface between the YBCO and the STO buffer, as shown in Figure 3(b) and (c), respectively. Modulations in the spacing between adjacent YBCO layers are also observed, as indicated by the arrow, and are attributed to the 3-dimensional growth mode of the YBCO, which is corroborated from AFM images of the YBCO surface shown in Figure 4. It should be noted that 3-dimensional growth is also typical of YBCO films grown by PLD and sputtering on single crystal oxide substrates [30].

XRD measurements (Bruker D8, Cu Kα) corroborate the single crystalline nature of the, *c*-axis oriented YBCO films. Figure 5 shows a ω-2θ survey scan of a 50-nm thick YBCO film grown on a 30 nm thick STO buffer. Only peaks associated with {00l} set of planes of YBCO and STO are observed, with the exception of a peak located at $2\theta \sim 44^0$ indicated by the asterisk; the origin of this peak will be discussed below. The out-of-plane lattice of our YBCO films grown on STO buffered Si(001) is $c = 11.66$ Å, which is comparable to the $c$ lattice constant of bulk YBCO. Rocking curve analysis about the YBCO (005) peak indicates a full-width-at-half-max (FWHM) of $\Delta\omega = 0.85^0$, as shown in the inset of Figure 5. To provide context, the rocking curve of a 50-nm thick YBCO film grown by PLD on a single crystal STO substrate is also shown in the inset of Figure 4. Though larger than the $\Delta\omega = 0.65^0$ FWHM of the YBCO film on STO single crystal, the FWHM of YBCO on the STO buffered Si remains comparable. The $\Delta\omega = 0.65^0$ in the YBCO on STO single crystal indicates mosaicity in the film, which is largely attributed to the difference in $c$ lattice constants between the STO ($c = 3.905$ Å) and the YBCO. More specifically, steps on the surface of the STO substrate are not commensurate with the YBCO unit-cell, which introduces dislocations in the film. Such mosaicity is further enhanced for YBCO grown on STO buffered Si since the STO on Si *per se* exhibits mosaicity, as evidenced by a FWHM $\Delta\omega = 0.40^0$ in the rocking curve of STO on Si. The mosaicity in the STO on Si is attributed to film relaxation, since the lattice constant of the former is larger than the surface lattice constant of the latter (3.84 Å), and the steps on the surface of the diamond cubic Si are not commensurate with the perovskite STO unit-cell [29,31]. Such mosaicity remains even after the formation of $SiO_x$ at the interface between the STO and Si.

Transport measurements indicate the YBCO films on STO buffered Si exhibit high $T_c$'s and narrow transition widths. Electrical transport measurements are performed using a standard ac lock-in technique on $1 \times 5$ mm samples in a 4-point geometry in which the current and voltage leads are in-line. Electrical contact is made to the YBCO by drawing conducting lines using Ag paint (SPI products) onto 15 nm thick Au contacts that are deposited onto the YBCO surface through a shadow mask by DC magnetron sputtering. Here we define $T_c$ as the temperature at which zero resistivity is observed in the transport characteristics. Figure 5 shows the resistivity versus temperature of a 50-nm thick YBCO film grown on 30 nm thick STO buffered Si. For comparison, the transport characteristics of a 50-nm thick YBCO film grown by PLD under equivalent conditions on a STO single crystal substrate are also shown. The $T_c$ of the YBCO on STO buffered Si is ~ 93 ± 0.5 K, which is higher than the $T_c$ of the YBCO on STO single crystal (~ 90 ± 0.5 K), as shown more clearly in the inset of Figure 6. The extrapolated zero temperature residual resistivity of the YBCO film grown on STO buffered Si is higher (~ 110 Ωcm ) than the YBCO on STO single crystal (~ 75 Ωcm), denoting a higher number of defects that give rise to elastic scattering, which would be consistent with the larger mosaicity in the former in comparison to the latter. Interestingly, the residual resistivity ratio (RRR), namely, the ratio between the resistivity at room temperature to the extrapolated resistivity at 0 K, is nearly a factor of 2 higher for the YBCO on STO buffered Si (RRR ~ 16), in comparison to the YBCO on STO single crystal (RRR ~ 9). The RRR is a function of temperature dependent inelastic scattering processes, such as phonons. In general, phonon transport in crystalline materials is higher than in amorphous materials. Though not fully understood, the enhanced RRR in the YBCO on STO buffered Si could be due to the layer of $SiO_x$ between the STO/YBCO and the Si, which acts as a "bottleneck" to phonons trying to escape as the sample is cooled.

The 93 K $T_c$ measured in our YBCO films is higher than the $T_c$'s reported thus far for YBCO grown on Si [9,11,20,32-36]. Furthermore, the thickness of our YBCO films (~50 nm) is also thinner than YBCO films previously grown on Si with or without a buffer. The higher robust $T_c$ achieved in such relatively thin films can be attributed to the single-crystalline nature of the STO buffer, which forms an epitaxial template for integrating YBCO on Si. Perhaps more intriguing is that the $T_c$ for the YBCO on STO buffered Si is higher than the $T_c$ for YBCO on STO single crystal. We attribute the enhanced $T_c$ to improved oxygenation of the YBCO on STO buffered Si in comparison to the YBCO on STO single crystal. First, the mosaicity of the former is larger, indicating a large number of dislocations which can facilitate oxygen flow through the film. Second, we note that an additional contributing factor to the higher $T_c$ may be the possible inclusion of Ag in the YBCO film, as indicated by the peak at $2\theta \sim 44^0$ in the XRD survey scan shown in Figure 5. Previous studies have shown that Ag inclusion also improves oxygenation of the YBCO [37-38]. Since Ag was not intentionally introduced in the polycrystalline PLD target, we suspect the Ag paste used to mount the STO buffered Si onto the heater plate is the origin of the XRD peak observed at $2\theta \sim 44^0$, as indicated by the asterisk [39-40].

Lastly, the single crystalline STO buffered Si forms a robust platform for integrating YBCO on Si. Despite extended exposure to ambient conditions and the absence of additional precautions protecting the 30-nm thick STO buffer on Si, the subsequent deposition of YBCO yields films that exhibit high $T_c$'s and narrow transition widths. Here, PLD is used to grow YBCO on $5 \times 5$ mm pieces of STO buffered Si. Though the technique of PLD can create high quality epitaxial YBCO films, the small size of the material plume created by the laser renders deposition on large substrates challenging. In this regard, the STO buffered Si platform is amenable to a variety of deposition techniques, such as sputtering, which can enable growth of YBCO on larger sized wafers.

4. **Conclusion**

In summary, we demonstrate an approach to epitaxially integrate YBCO films on Si (001) through the use of a 30-nm thick single crystalline STO buffer grown by MBE. X-ray diffraction and cross-sectional scanning transmission electron microscopy confirm the single crystalline nature of the YBCO and STO buffer. Our 50-nm thick YBCO films on Si exhibit transition temperatures of ~ 93 K, and narrow transition widths (< 5 K). The realization of epitaxial YBCO films with high transition temperatures on Si opens the pathway to explore YBCO for various applications.

## Acknowledgments

This work was supported by the University of Texas Arlington, NSERC, and by the National Science Foundation (NSF) under award No. DMR-1508530.

**Figures**

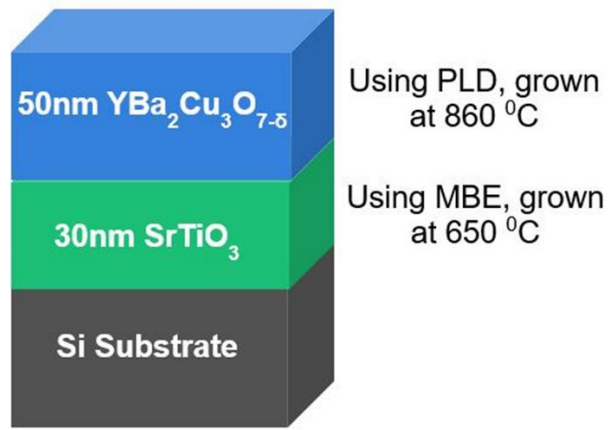

**Figure 1**. Schematic of heterostructure grown by PLD and MBE. The 30nm STO is grown on Si (001) using oxide molecular beam epitaxy (MBE), and the 50nm YBCO is grown using pulsed laser deposition (PLD).

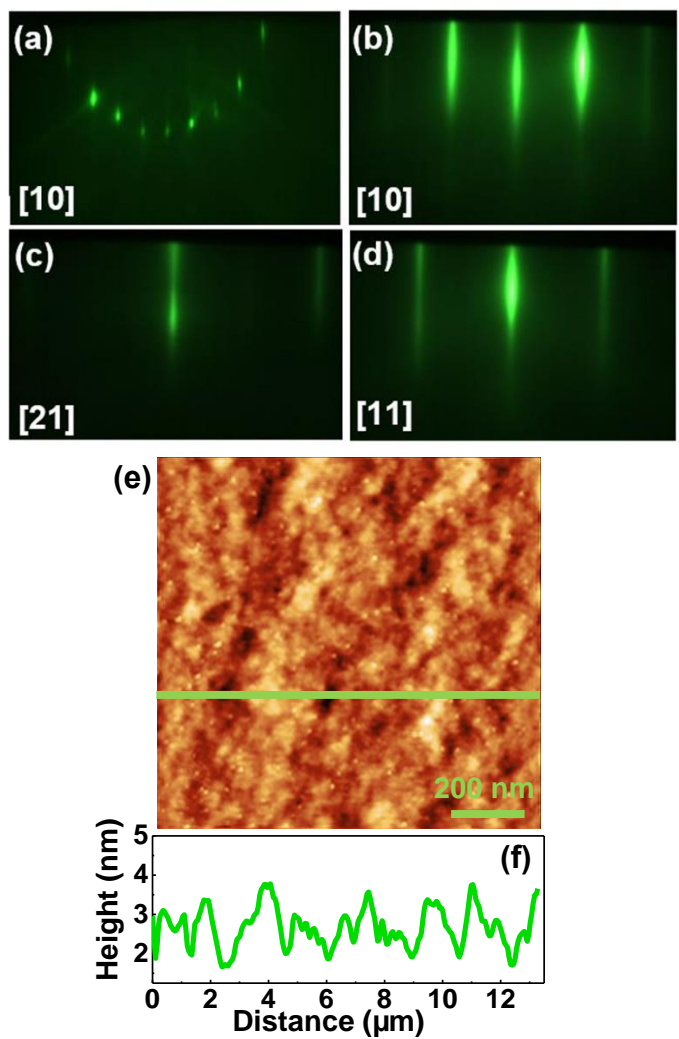

**Figure 2.** (a) RHEED patterns of clean Si along [10] direction; (b) After 30 nm STO grown epitaxially on Si (001) taken along [10] and (c) [21] and (d) [11] directions. (e) AFM image of 30 nm STO on Si (001). The surface morphology confirms 2-dimenssional growth. (f) Line-profile analysis of the surface roughness taken along the line indicated in (e).

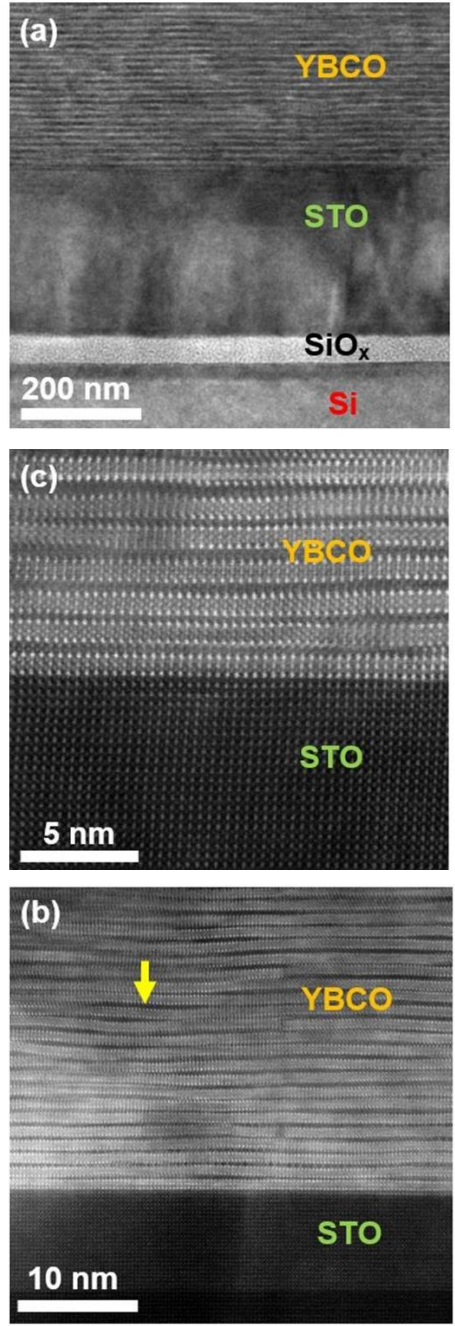

**Figure 3**. (a) HRTEM image of the YBCO/STO/Si heterostructure. A layer of $SiO_x$ forms between the STO and Si during the deposition of YBCO. (b) STEM image of YBCO on STO buffer. Modulations in the spacing between layers of YBCO are observed, as indicated by the arrow. (c) STEM image of the atomically abrupt interface between the YBCO and STO buffer.

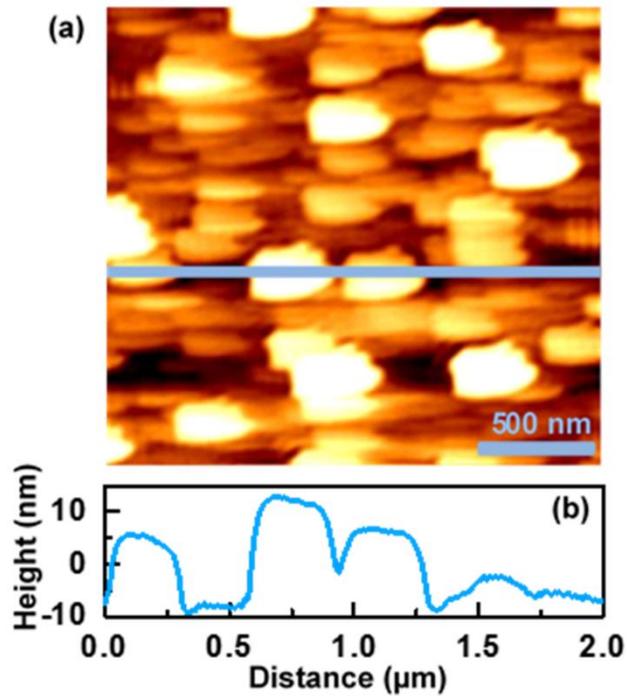

**Figure 4.** (a) AFM image of a 50 nm thick YBCO film grown on STO buffered Si(001). The surface morphology indicates a 3-dimensional growth mode. (b) Line-profile analysis of the surface roughness taken along the line indicated in (a).

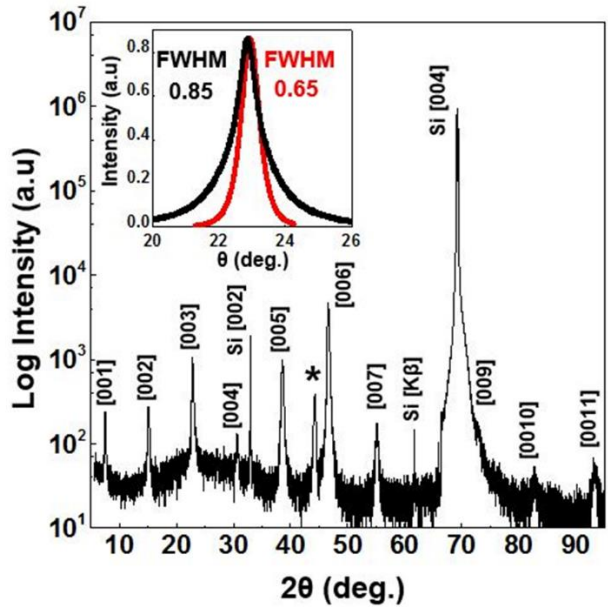

**Figure 5**. XRD of a YBCO/STO/Si heterostructure. The ω-2θ survey shows peaks associated with the {00l} planes of YBCO and STO. Rocking curve of the YBCO grown on STO buffered Si (001) is shown in the inset (black). For comparison, the rocking curve of a 50-nm thick YBCO film grown on STO single crystal substrate is also shown (red). A peak believed to arise from Ag is also observed at 2θ ~ $44^0$, as indicated by the asterisk.

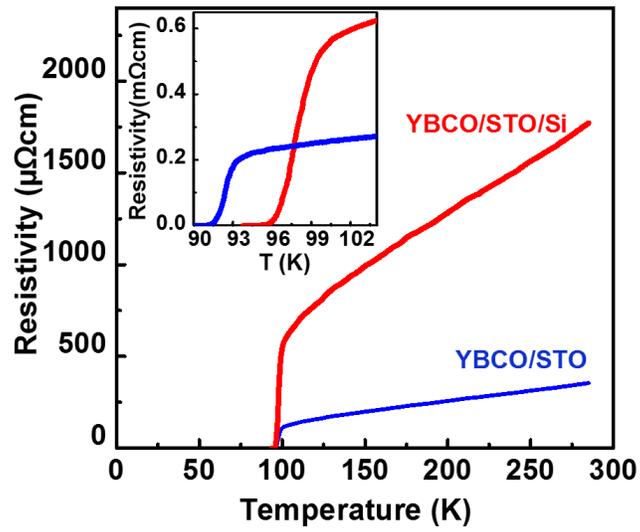

**Figure 6.** Transport characterization of YBCO/STO/Si (red) and YBCO/STO (blue) heterostructures. The 50 nm thick YBCO grown on STO buffered Si (001) exhibits a $T_c \sim 93$ K, as shown more clearly in the inset.